\documentclass[aps, showpacs, showkeys,nofootinbib,floatfix]{revtex4}

\usepackage{amssymb}
\usepackage{amsmath}
\usepackage{graphicx}

\begin{document}

\title{Observational constraint on generalized Chaplygin gas model}

\author{Jianbo Lu}
\email{lvjianbo819@163.com}
\author{Yuanxing Gui}
\author{Lixin Xu}

\affiliation{School of Physics and Optoelectronic Technology, Dalian
University of Technology, Dalian, 116024, P. R. China}

\begin{abstract}

We investigate observational constraints on  the generalized
Chaplygin gas (GCG) model as the
 unification of dark matter and dark energy from  the latest
observational data: the Union SNe Ia data, the  observational Hubble
data, the SDSS baryon acoustic peak and the five-year WMAP shift
parameter. It is obtained  that the best fit values of the GCG model
parameters with their confidence level are
$A_{s}=0.73^{+0.06}_{-0.06}$ ($1\sigma$) $^{+0.09}_{-0.09}$
$(2\sigma)$, $\alpha=-0.09^{+0.15}_{-0.12}$ ($1\sigma$)
$^{+0.26}_{-0.19}$ $(2\sigma)$. Furthermore in this model, we can
see that the evolution of equation of state (EOS) for dark energy is
similar to quiessence, and its current best-fit value is
$w_{0de}=-0.96$ with the $1\sigma$ confidence level $-0.91\geq
w_{0de}\geq-1.00$.

\end{abstract}


\maketitle

\section{$\text{Introduction}$}

 {\small {~~~}}~ The recently cosmic observations from the
type Ia
 supernovae (SNe Ia) \cite{SNe}, the cosmic microwave background (CMB)
\cite{CMB}, the clusters of galaxies \cite{LSS} etc., all suggest
that the expansion of present universe is speeding up rather than
slowing down.  And it indicates that baryon matter component is
about 5\% for total energy density, and about 95\% energy density in
universe is invisible.  Considering the four-dimensional  standard
cosmology, the accelerated expansion of the present universe is
usually attributed to the fact that dark energy (DE) is an exotic
component with negative pressure. And it is shown that DE takes up
about two-thirds of the total energy density from cosmic
observations. Many kinds of DE models have already been constructed
such as $\Lambda$CDM \cite{LCDM}, quintessence \cite{quintessence},
phantom \cite{phantom}, quintom \cite{quintom}, generalized
Chaplygin gas (GCG) \cite{GCG}, modified Chaplygin gas \cite{MCG},
holographic dark energy \cite{holographic}, agegraphic dark
energy\cite{agegraphic}, and so forth. Furthermore,
model-independent method\footnote{Using mathematical fundament, one
expands  equation of state  of DE $w_{de}$ or deceleration parameter
$q$ with respect to scale factor $a$ or redshit $z$. For example,
$w_{de}(z)=w_{0}$=const \cite{independent1},
$w_{de}(z)=w_{0}+w_{1}z$\cite {independent2},
$w_{de}(z)=w_{0}+w_{1}\ln(1+z)$ \cite{independent3},
$w_{de}(z)=w_{0}+\frac{w_{1}z}{1+z}$ \cite{independent4},
$q(z)=q_{0}+q_{1}z$ \cite{independent1}, $q(z)=q_{0}+
\frac{q_{1}z}{1+z}$ \cite{independent5}, where $w_{0}$, $w_{1}$, or
$q_{0}$, $q_{1}$ are model parameters.} and modified gravity
theories (such as scalar-tensor cosmology \cite{scalar}, braneworld
models \cite{braneworld})  to interpret accelerating universe have
also been discussed.

It is well known that the GCG model have been widely studied  for
interpreting the accelerating universe \cite{GCGpapers}. The most
interesting property for this scenario is that, two unknown dark
sections in universe--dark energy and dark matter  can be unified by
using an exotic equation of state (EOS). In this paper, we use the
latest
 observational data: the Union SNe Ia data \cite{307Union},
the observational Hubble data (OHD) \cite{OHD},  the baryon acoustic
oscillation (BAO) peak from Sloan Digital Sky Survey (SDSS)
\cite{SDSS} and the five-year WMAP CMB shift parameter \cite{5yWMAP}
to constrain the GCG model. And we  discuss whether the parameter
degeneration \cite{GCG2}\cite{GCG3} for the GCG model can be broken
by the latest observed data,  since  it is always expected that the
model degeneration problem can be solved  by the more accurate
observational data.

The paper is organized as follows. In section 2, the GCG model as
the unification of dark matter and dark energy is introduced
briefly. Based on the   observational data, we constrain the GCG
model parameter in section 3.  The evolutions of  EOS of DE and
deceleration parameter for GCG model are presented in section 4.
Section 5 is the conclusions.

\section{$\text{generalized Chaplygin gas model}$}

 The GCG background fluid with its energy density
$\rho_{GCG}$ and pressure $p_{GCG}$ are related by the EOS
\cite{GCG}
\begin{equation}
p=-\frac{A}{\rho^{\alpha}},\label{e1}
\end{equation}
where $A$  and $\alpha$ are parameters in the model. When $\alpha =
1$, it is reduced to the CG scenario.

Considering the Friedmann-Robertson-Walker (FRW) cosmology, by using
the energy conservation equation: $d(\rho a^{3})=-pd(a^{3})$, the
energy density of GCG can be derived as
\begin{equation}
\rho _{GCG}=\rho _{0GCG}[A_s+(1-A_s)(1+z)^{3(1+\alpha )}]^{\frac
1{1+\alpha }},\label{e2}
\end{equation}
where $a$ is the scale factor, A$_s=\frac A{\rho _0^{1+\alpha }}$.
For the GCG model, as a scenario of the unification of  dark matter
and dark energy, the GCG fluid is decomposed into two components:
the dark energy component and the dark matter component, i.e., $
\rho _{GCG}=\rho _{de}+\rho _{dm}$, $p_{GCG}=p_{de}$. Then according
to the general recognition about dark matter
\begin{equation}
\rho _{dm}=\rho _{0dm}(1+z)^3,\label{e3}
\end{equation}
the energy density of the DE in the GCG model is given by
\begin{equation}
\rho _{de}=\rho_{GCG} -\rho _{dm}=\rho
_{0GCG}[A_s+(1-A_s)(1+z)^{3(1+\alpha )}]^{\frac 1{1+\alpha }}-\rho
_{0dm}(1+z)^3.\label{e4}
\end{equation}

Next, we assume the universe is filled with two components, one is
the GCG component, and the other is baryon matter component, ie., $
\rho _t=\rho _{GCG}+\rho _b.$ In a flat universe, making use of the
Friedmann equation, the Hubble parameter $H$ is expressed as
\begin{equation}
H^2=\frac{8\pi G \rho_{t}}{3}=H_0^2E^{2}=H_0^2\{(1-\Omega
_{0b})[A_s+(1-A_s)(1+z)^{3(1+\alpha )}]^{\frac 1{1+\alpha }}+\Omega
_{0b}(1+z)^3\}.\label{e5}
\end{equation}
Where  $H_{0}=100h~km~S^{-1}Mpc^{-1}$ is the present Hubble
constant,  $h = 0.72 \pm 0.08$ is given by Hubble Space Telescope
key projects \cite{Hprior}. $\Omega_{0b}$ is the present value of
dimensionless baryon matter density, and a joint analysis of
five-year WMPA, SNe Ia and BAO data gives $\Omega_{0b}h^{2}=
0.02265\pm 0.00059$
 \cite{5yWMAPSNeBAO}. In the following section, we will use the cosmic observations to constrain
the GCG model parameter ($A_{s},\alpha$).
\\

\section{$\text{ Constraint on GCG model parameter}$}

It is necessary for the investigation of type Ia supernovae to
explore dark energy and constrain the models. Since SNe Ia behave as
excellent standard candles, they can be used to directly measure the
expansion rate of the universe up to high redshift with comparing
with the present rate. Theoretical dark-energy model parameters are
determined by minimizing the quantity \cite{chi2SNe}
\begin{equation}
\chi^{2}_{SNe}(\theta)=\sum_{i=1}^{N}\frac{(\mu_{obs}(z_{i})
-\mu_{th}(\theta;z_{i}))^2}{\sigma^2_{obs;i}},\label{e6}
\end{equation}
where $N=307$ for the Union SNe Ia data \cite{307Union}, which
includes the SNe samples from the Supernova Legacy Survey
\cite{SNLS}, ESSENCE Surveys \cite{ESSENCE}, distant SNe discovered
by the Hubble Space Telescope  \cite{HST}, nearby SNe \cite{nearby}
 and several other, small
data sets. The $1\sigma$ error $\sigma_{obs;i}$ are from flux
uncertainty, intrinsic dispersion of SNe Ia absolute magnitude and
peculiar velocity dispersion, which are assumed to be Gaussian and
uncorrelated.  $\theta$ denotes the model parameters. $\mu_{obs}$ is
the observed value of distance modulus and can be given by the SNe
dataset. The theoretical distance modulus $\mu_{th}$ is defined as
\begin{equation}
\mu_{th}(z_{i})\equiv m_{th}(z_{i})-M.\label{e7}
\end{equation}
Here $m_{th}(z)$ is the apparent magnitude of the SNe at peak
brightness
\begin{equation}
m_{th}(z)=\overline{M}+5log_{10}(D_{L}(z)),\label{e8}
\end{equation}
and absolute magnitude $M$ can be given by relating to the magnitude
zero point offset $\overline{M}$,
\begin{equation}
\overline{M}=M+\mu_{0}\label{e9}
\end{equation}
with $\mu_{0}=
5log_{10}(\frac{H_{0}^{-1}}{Mpc})+25=42.38-5log_{10}h$. Thus
according to  Eqs. (\ref{e7}), (\ref{e8}) and (\ref{e9}), the
theoretical distance modulus can be written as
\begin{equation}
\mu_{th}(z) =5log_{10}(D_{L}(z))+\mu_{0},\label{e10}
\end{equation}
where $D_{L}(z)$ is the Hubble free  luminosity distance
\begin{equation}
D_{L}(z)=H_{0}d_{L}(z)=(1+z)\int_{0}^{z}\frac{dz^{'}}{E(\theta;z^{'})}.\label{e11}
\end{equation}
Since the nuisance parameter $\mu_{0}$ is independent of the data
and the dataset, from above equations one can see that the distance
modulus of different SNe (i.e. at different redshift $z$),
$\mu(z_{i})$ and $\mu(z_{j})$ are uncorrelated. So, the covariance
matrix included in the $\chi^{2}_{SNe}$ (Eq. (\ref{e6}))  is
diagonal with entries $\sigma_{i}$.

Furthermore, by expanding the $\chi^{2}_{SNe}$ of expression
(\ref{e6}) relative to $\mu_{0}$, the minimization with respect to
$\mu_{0}$ can be made trivially
\cite{chi2SNe}\cite{SNeABC}\cite{chi2sneli}
\begin{equation}
\chi^{2}_{SNe}(\theta)=A(\theta)-2\mu_{0}B(\theta)+\mu_{0}^{2}C,\label{e12}
\end{equation}
where
\begin{equation}
A(\theta)=\sum_{i=1}^{N}\frac{[\mu_{obs}(z_{i})-\mu_{th}(z_{i};\mu_{0}=0,\theta)]^{2}}{\sigma_{i}^{2}},\label{e13}
\end{equation}
\begin{equation}
B(\theta)=\sum_{i=1}^{N}\frac{\mu_{obs}(z_{i})-\mu_{th}(z_{i};\mu_{0}=0,\theta)}{\sigma_{i}^{2}},\label{e14}
\end{equation}
\begin{equation}
C=\sum_{i=1}^{N}\frac{1}{\sigma_{i}^{2}}.\label{e15}
\end{equation}
Evidently, Eq. (\ref{e6}) has a minimum for $\mu_{0}=B/C$ at
\begin{equation}
\widetilde{\chi}^{2}_{SNe}(\theta)=A(\theta)-B(\theta)^{2}/C.\label{e16}
\end{equation}
 Since $\chi^{2}_{SNe,min}=\widetilde{\chi}^{2}_{SNe,min}$ and $\widetilde{\chi}^{2}_{SNe}$ is independent of nuisance
parameter $\mu_{0}$  \cite{chi2sneli}, here we utilize
  expression (\ref{e16}) to displace (\ref{e6}) for SNe
constraint.

\begin{table}
\vspace*{-12pt}
\begin{center}
\begin{tabular}{c | c  }
\hline\hline $z$
&~0.09~~~~~~0.17~~~~~~0.27~~~~~~0.40~~~~~~0.88~~~~~~1.30~~~~~~1.43~~~~~~1.53~~~~~~1.75~~~~~~
\\\hline
   $H(z)$ (kms$^{-1}$~Mpc)$^{-1}$   &~~~69~~~~~~~~83~~~~~~~70~~~~~~~~~~87~~~~~~~117~~~~~~~168~~~~~~~177~~~~~~~140~~~~~~202~~~~~~
 \\\hline
    $1\sigma$ uncertainty
    &~~$\pm$12~~~~~$\pm$8.3~~~~~~$\pm$14~~~~$\pm$17.4~~~~$\pm$23.4~~~~$\pm$13.4~~~~$\pm$14.2~~~~~$\pm$14~~~~$\pm$40.4~~~~~~
  \\\hline\hline
       \end{tabular}
       \end{center}
     Table 1. The observational $H(z)$ data \cite{OHD}\cite{OHDdata}.
       \end{table}

 Since the Hubble parameter $H(z)$ depends on the differential
age of the universe,
\begin{equation}
H(z)=-\frac{1}{1+z}\frac{dz}{dt}.\label{e17}
\end{equation}
the value of $H(z)$ can directly be measured through a determination
of $dz/dt$. By using the differential ages of passively evolving
galaxies from the GDDS \cite{Hdata1} and archival data
\cite{Hdata2}, Ref.  \cite{OHD} got nine values of $H(z)$ in the
range of $0<z<1.8$ (see Table 1). Here the observed Hubble data
$H(z_{i})$ and $H(z_{j})$ are uncorrelated, for they are obtained by
the observations of galaxies at different redshift. Using these nine
observational Hubble data one can constrain DE models by minimizing
\cite{OHDdata}\cite{chi2OHD}
\begin{equation}
\chi^2_{Hub}(H_{0},
\theta)=\sum_{i=1}^{N}\frac{\left[H_{th}(H_{0},\theta,z_i)-H_{obs}(
z_i)\right]^2}{\sigma^2_{obs;i}},\label{e18}
\end{equation}
where $H_{th}$ is the predicted value of the Hubble parameter,
$H_{obs}$ is the observed value, $\sigma_{obs;i}$ is the $1\sigma$
uncertainty of the  measurement of standard deviation. Here $H_{0}$
contained in the $\chi^{2}_{Hub}(H_{0},\theta)$ as a nuisance
parameter is marginalized by integrating
 the likelihood
$L(\theta) =\int d H_{0}P(H_{0})\exp$
 $(-\chi^{2}(H_{0},\theta)/2)$.
$P(H_{0})$ is the prior distribution function of the present Hubble
constant, and a Gaussian prior $H_{0}= 72 \pm 8~km~S^{-1}Mpc^{-1}$
\cite{Hprior} is adopted in this paper.

Using a joint analysis of Union SNe Ia data and OHD (i.e.,
$\chi_{total}^{2}=\chi_{SNe}^{2}+\chi_{Hub}^{2}$),  Fig. 1 shows the
constraint on GCG parameter space $A_{s}$-$\alpha$ at  the $1\sigma$
(68.3\%) and $2\sigma$ (95.4\%) confidence levels. For this analysis
the best fit parameters are $A_{s}=0.80$ and $\alpha=0.42$. It is
obvious  that two model parameters, $A_{s}$ and $\alpha$, are
degenerate. And it can be seen that
 model parameter $\alpha$  has the lager variable range. Then in order
 to get the stringent constraint
 and  diminish systematic uncertainties, in what follows we  combine  the standard ruler data (the BAO peak
from SDSS and the five-year WMAP CMB shift parameter $R$) with the
Union SNe Ia data and the OHD  to constrain the GCG model.

\begin{figure}[!htbp]
   \includegraphics[width=6cm]{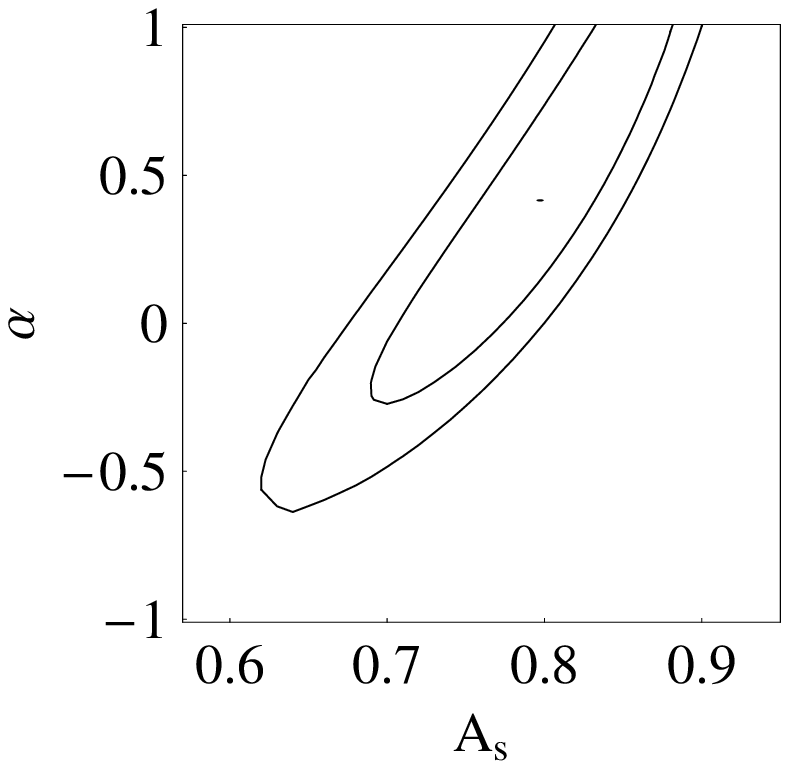}\\
Fig. 1. The 68.3\%  and 95.4\% confidence level contours for $A_{s}$
versus $\alpha$ from the Union SNe data plus the OHD.
\end{figure}

Because the universe has a fraction of baryon, the acoustic
oscillations in the relativistic plasma would be imprinted onto the
late-time power spectrum of the nonrelativistic matter
\cite{27Eisenstein}. Then the observations of acoustic signatures in
the large-scale clustering of galaxies are very important for
constraining cosmological models. From the BAO constraint, the best
fit values of parameters in the DE models can be determined by
constructing \cite{chi2bao}
\begin{equation}
\chi^{2}_{BAO}(\theta)=\frac{[A(\theta)-A_{obs}]^{2}}{\sigma_{A}^{2}}.\label{e19}
\end{equation}
Where
\begin{equation}
A(\theta)=\sqrt{\Omega_{0m}}E(z_{BAO})^{-1/3}[\frac{1}{z_{BAO}}\int_{0}^{z}\frac{dz^{'}}{E(z^{'};\theta)}]^{2/3},\label{e20}
\end{equation}
$\Omega_{0m}$ is the effective matter density parameter given by
$\Omega_{0m}=\Omega_{0b}+(1-\Omega_{0b})(1-A_{s})^{\frac{1}{1+\alpha}}$
\cite{GCG2}\cite{GCG3}\cite{GCG1}. The observed value  $A_{obs}$
with its $1\sigma$ error $\sigma_{A}$ is
$A_{obs}=0.469(n_{s}/0.98)^{-0.35}\pm0.017$ measured from the SDSS
at $z_{BAO}=0.35$, here $n_{s}$ is the scalar spectral index
\cite{scalarspectral} and its value is taken to be 0.96 as shown in
Ref. \cite{5yWMAPSNeBAO}.

The structure of the anisotropies of the cosmic microwave background
radiation depends on two eras in cosmology, i.e., the last
scattering era and today. They can also be applied to limit  DE
models by minimizing \cite{chi2CMB}
\begin{equation}
\chi^{2}_{CMB}(\theta)=\frac{(R(\theta)-R_{obs})^{2}}{\sigma_{R}^{2}}.\label{e21}
\end{equation}
Where the shift parameter \cite{RCMB}
\begin{equation}
R(\theta)=\sqrt{\Omega_{0m}}\int_{0}^{z_{rec}}\frac{dz^{'}}{E(z^{'};\theta)},\label{e22}
\end{equation}
 $z_{rec}=1089$ is the redshift of recombination. The
observed value  $R_{obs} = 1.710$, and its corresponding $1\sigma$
error is $\sigma_{R}=0.019$  according to the five-year WMAP result
\cite{5yWMAP}.

\begin{figure}[!htbp]
   \includegraphics[width=6cm]{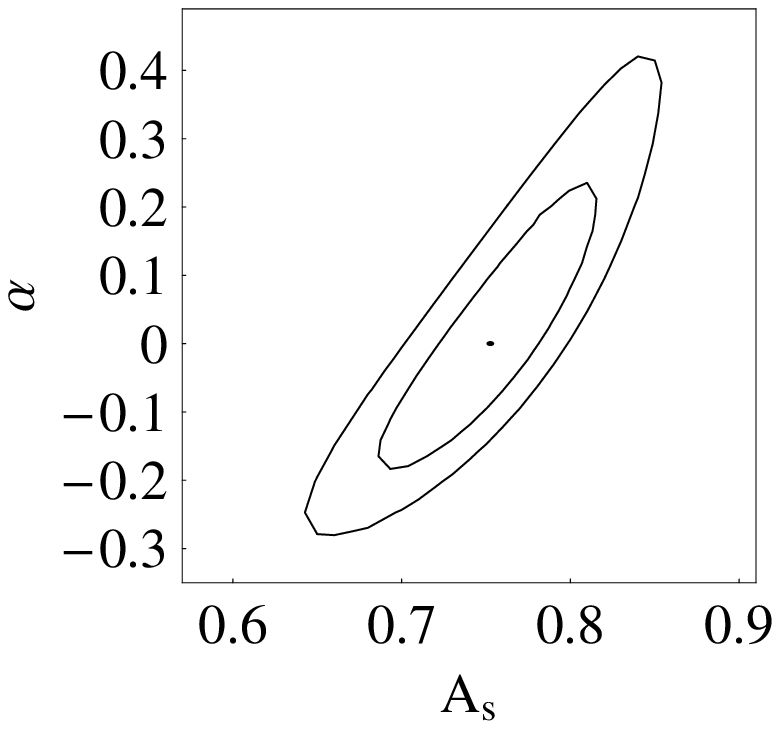}~~~~~~~
   \includegraphics[width=6cm]{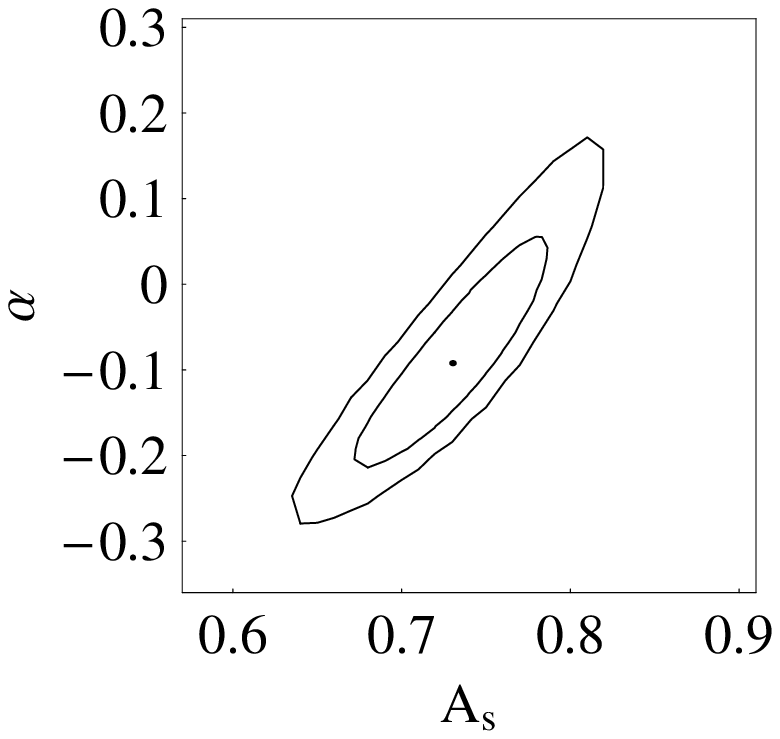}\\
~~~~~~~~~~~~~~~~~~~~~~~~~~~~~~~(a)~~~~~~~~~~~~~~~~~~~~~~~~~~~~~~~~~~~~~~~~~~~~~~~~~~~~~~~~~~~~(b)~~~~~~~~~~~~~~~~~~~~\\
Fig. 2. The 68.3\%  and 95.4\% confidence level contours for $A_{s}$
versus $\alpha$ from the Union SNe data plus the BAO data (a), and a
combined analysis of the Union SNe,  OHD,   BAO and  CMB data (b).
\end{figure}

Above four observational data are uncorrelated for each other, since
they are given by different experiments and methods. Then the total
likelihood $\chi^{2}_{total}$ can be constructed as
\begin{equation}
\chi^{2}_{total}=\chi^{2}_{SNe}+\chi^{2}_{Hub}+\chi^{2}_{BAO}+\chi^{2}_{CMB}.\label{e23}
\end{equation}
Using Eq. (\ref{e23}) we get the best fit values of GCG model
 parameters  $(A_{s},\alpha)$ are (0.73,-0.09) with
$\chi^{2}_{min}=322.87$, and the reduced $\chi^{2}$ value
is\footnote{The value of dof (degrees of freedom) for the model
equals the number of observational data points minus the number of
parameters.} $\chi^{2}_{min}$/dof=1.03.  The $1\sigma$   and
$2\sigma$  confidence level contours of GCG model parameters are
plotted in Fig. 2 (b). From this figure, we obtain the values of
model parameters with the  confidence levels,
$A_{s}=0.73^{+0.06}_{-0.06}$ ($1\sigma$) $^{+0.09}_{-0.09}$
$(2\sigma)$ and $\alpha=-0.09^{+0.15}_{-0.12}$ ($1\sigma$)
$^{+0.26}_{-0.19}$ $(2\sigma)$. It can be seen that  parameters
$A_{s}$ and $\alpha$ are also degenerate, and at the $1\sigma$
confidence level  these results are consistent with the standard
dark energy plus dark matter scenario (i.e., the case of
$\alpha=0$).   Furthermore, one can see that this constraint on
parameter $\alpha$ is more stringent than the results in Refs.
\cite{GCG2}\cite{GCG3}, where the constraint results for the GCG
model parameters are $A_{s}=0.70^{+0.16}_{-0.17}$ and
$\alpha=-0.09^{+0.54}_{-0.33}$ at $2\sigma$ confidence level
  by using the X-ray gas mass fractions of galaxy
clusters and the dimensionless coordinate distance of SNe Ia and
FRIIb radio galaxies \cite{GCG2}, and $A_{s}=0.75^{+0.08}_{-0.08}$,
$\alpha=0.05^{+0.37}_{-0.26}$ at  $2\sigma$ confidence level
 by means of the observational Hubble data, the 115 SNLS
SNe Ia data and the SDSS baryonic acoustic oscillations peak
\cite{GCG3}.

 At last, we  also consider the constraint on the GCG
model parameter from  a combination of
 Union SNe Ia and BAO
data, the best fit  happens at $A_{s}=0.75$ and $\alpha=0$, which
can be reduced to the standard dark energy plus dark matter
scenario. But at their confidence levels, the two parameters are
also highly degenerate. In Fig. 2 (a), we display the constraint
result for this analysis.

\section{$\text{Constraint on EOS of dark energy and deceleration
parameter}$}

\begin{figure}[!htbp]
\includegraphics[width=6cm]{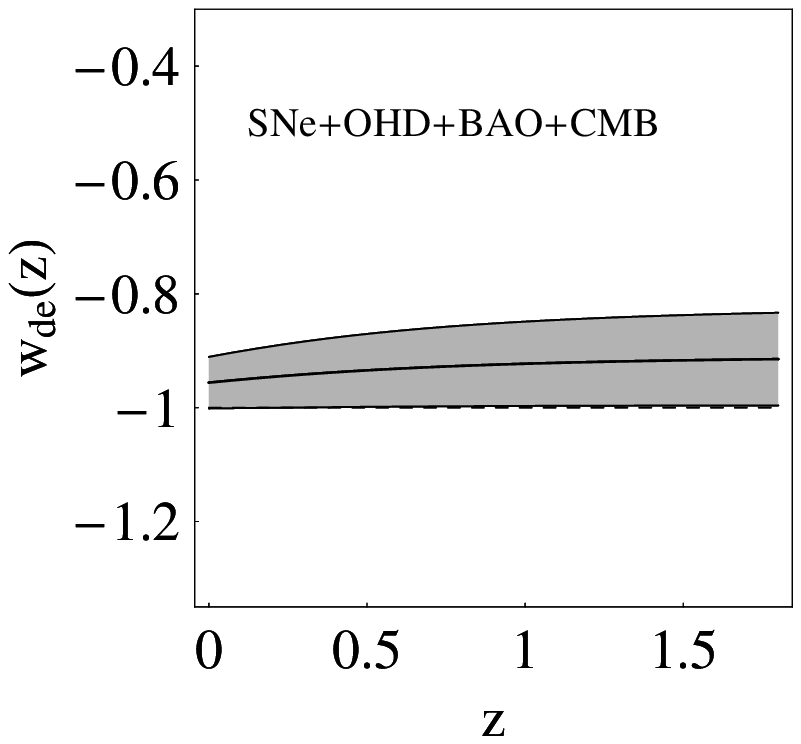}
~~~~~~~\includegraphics[width=6cm]{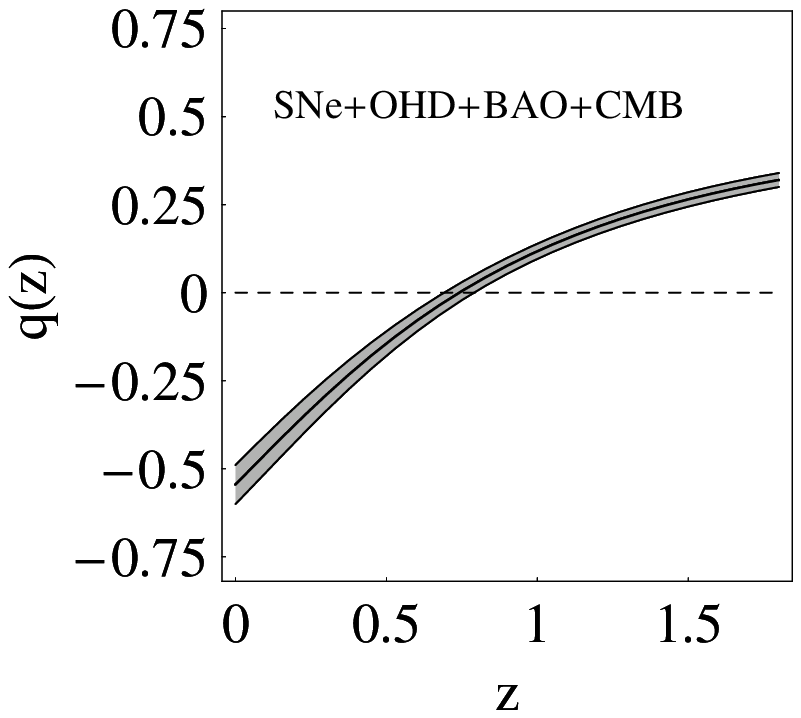}\\
 ~~~~~~~~~~~~~~~~~~~~~~~~~~~~~~~(a)~~~~~~~~~~~~~~~~~~~~~~~~~~~~~~~~~~~~~~~~~~~~~~~~~~~~~~~~~~~~(b)~~~~~~~~~~~~~~~~~\\
Fig. 3. The best fits of $w_{de}(z)$ and $q(z)$ with $1\sigma$
confidence level for GCG model.
\end{figure}

 According the Eq. (\ref{e5}), deceleration parameter $q$ in the GCG model can be
obtained by
\begin{equation}
q=(1+z)\frac{1}{H}\frac{dH}{dz}-1.\label{e24}
\end{equation}
The equation of state  of dark energy is derived as
\begin{equation}
w _{de}=\frac{p_{de}}{\rho_{de}}=\frac{-(1-\Omega
_{0b})A_s[A_s+(1-A_s)(1+z)^{3(1+\alpha )}]^{-\frac \alpha {1+\alpha
}}}{ (1-\Omega _{0b})[A_s+(1-A_s)(1+z)^{3(1+\alpha )}]^{\frac
1{1+\alpha }}-\Omega _{0dm}(1+z)^3},\label{e25}
\end{equation}
where $\Omega_{0dm}$ is the present value of dimensionless dark
matter density. Based on Eqs. (\ref{e24}) and (\ref{e25}), the
confidence levels of the best fit $w_{de}(z)$ and $q(z)$ calculated
by using
 the covariance matrix are plotted in Fig. 3.
 From Fig. 3 (a), it is easy to see that
 the  best fit value $w_{0de} \equiv w_{de}(z=0)=-0.96>-1$, and the
 $1\sigma$
confidence level of $w_{0de}$ is $-0.91\geq w_{0de}\geq-1.00$.
 In addition, it can be found that the best fit evolution of $w_{de}(z)$ for GCG
 is similar to the quiessence model ($w_{de}(z)$ = const $\neq -1$).
 From
Fig. 3 (b), we can see that the best fit values of transition
redshift and current deceleration parameter with confidence levels
are $z_{T}=0.74^{+0.04}_{-0.05}$ $(1\sigma)$,
$q_{0}=-0.55^{+0.05}_{-0.06}$ $(1\sigma)$. One knows that  $z_{T}$
describes the  expansion of universe from deceleration to
acceleration, and $q_{0}$ indicates the expansion rhythm of current
universe. Comparing our results with Ref. \cite{zTqSNe}, where
$z_{T}=0.49^{+0.14}_{-0.07}$ $(1\sigma)$ and
$q_{0}=-0.73^{+0.21}_{-0.20}$ $(1\sigma)$ are obtained from Union
SNe Ia data by using a linear two-parameter expansion for the
decelerating parameter, $q(z)=q_{0}+q_{1}z$, it is clear
 for our constraint,  that the universe  tends to an earlier
time to acceleration and a milder  expansion rhythm at present.

\section{$\text{Conclusion}$}

The constraints on the GCG model as the
 unification of dark matter and dark energy are studied in this paper by using
  the latest observational data: the Union SNe Ia data, the
observational Hubble data, the SDSS baryon acoustic peak and the
five-year WMAP shift parameter. We find that the model parameters
$A_{s}$ and $\alpha$ are degenerate, and their values  are
constrained to $A_{s}=0.73^{+0.06}_{-0.06}$ ($1\sigma$)
$^{+0.09}_{-0.09}$ $(2\sigma)$ and $\alpha=-0.09^{+0.15}_{-0.12}$
($1\sigma$) $^{+0.26}_{-0.19}$ $(2\sigma)$. This constraint on
parameter $\alpha$ is more stringent than the results in Refs.
\cite{GCG2}\cite{GCG3}. Furthermore, it is shown that the evolution
of EOS of dark energy for the GCG model is similar to quiessence,
and the best fit value of current EOS of DE $w_{0de}=-0.96>-1$. And
it indicates that the values of transition redshift and current
deceleration parameter are $z_{T}=0.74^{+0.05}_{-0.05}$ $(1\sigma)$,
$q_{0}=-0.55^{+0.06}_{-0.05}$ $(1\sigma)$.

\textbf{\ Acknowledgments } The research work is supported by  NSF
(10573004) of PR China.

\end{document}